\documentclass[prl,twocolumn,showpacs,preprintnumbers,amsmath,amssymb]{revtex4}
\usepackage{amsfonts}
\usepackage{graphics}
\usepackage{rotating}
\usepackage{makeidx}
\usepackage{amsthm}
\input xy
\xyoption{all}
\newcommand{\ket}[1]{\left| #1 \right\rangle}
\newcommand{\bra}[1]{\left\langle #1 \right|}
\newcommand{\abs}[1]{\left| #1 \right|}

\newcommand{\inprod}[2]{\left\langle \left. #1 \, \right| #2 \right\rangle}

\newtheorem{lemma}{Lemma}
\newtheorem{proposition}{Proposition}

\DeclareMathOperator{\tr}{tr}

\DeclareMathOperator{\PTr}{PTr}

\begin{document}

\title{The Parts Determine the Whole except for
$n$-Qubit Greenberger-Horne-Zeilinger States}

\author{Scott N. Walck}
  \email{walck@lvc.edu}
\author{David W. Lyons}
  \email{lyons@lvc.edu}
\affiliation{Lebanon Valley College, Annville, PA 17003}

\date{August 5, 2008}

\begin{abstract}
The generalized $n$-qubit Greenberger-Horne-Zeilinger (GHZ) states
and their local unitary equivalents are the only pure states
of $n$ qubits that are not uniquely determined
(among arbitrary states, pure or mixed)
by their reduced density matrices of $n-1$ qubits.
Thus, the generalized GHZ states are the only ones containing
information at the $n$-party level.
\end{abstract}

\pacs{03.67.Mn,03.65.Ta,03.65.Ud}

\maketitle

There are a number of perspectives from which to
attempt an understanding of multiparty quantum entanglement.
Entanglement can be viewed
in terms of its ability to reject local
realism and local hidden variable theories
\cite{bell64,greenberger89}.
From an operational point of view, entanglement can be
classified based on local operations
and classical communication (LOCC) and related notions
\cite{nielsen99,dur00}.
Quantum entanglement can be viewed as a resource for
quantum information technologies
\cite{bennett93,shor94}.
Finally, entanglement
can be viewed in terms of the
relationship between states of a quantum system and its subsystems
\cite{linden02,linden02b,jones05}.
Partial progress has been made in each of these perspectives,
but none has emerged as a definitive way to think about
entanglement.
This paper deals with the last, ``parts and whole'' perspective
on quantum entanglement, and makes a connection
with the first, ``rejection of local realism'' view.

The ``parts and whole'' view of quantum entanglement
asks to what extent a quantum state can be described
by the (typically mixed) states of its subsystems.
In particular, can an unknown state be uniquely determined,
or identified, by giving the states of its subsystems
(its reduced density matrices)?
In \cite{linden02,linden02b},
Linden, Popescu, and Wootters proved the surprising result
that almost all $n$-party pure states are determined
by their reduced density matrices (RDMs).
From the ``parts and whole'' perspective,
the most entangled states
are those that remain undetermined even after giving every
possible reduced density matrix for every subsystem.
In \cite{walcklyons08}, the present authors proved
that the only $n$-qubit states undetermined
by their reduced density matrices
\emph{among other pure states}
are the generalized $n$-qubit
Greenberger-Horne-Zeilinger (GHZ) states,
\[
\alpha \ket{00 \cdots 0} + \beta \ket{11 \cdots 1} ,
 \hspace{1cm} \alpha \beta \neq 0
\]
and their local unitary (LU) equivalents.

In this Letter, we prove the stronger result
that the generalized $n$-qubit GHZ states and their
LU equivalents are precisely the states that are
undetermined by their reduced density matrices,
among arbitrary states (pure or mixed).
From the ``parts and whole'' perspective on quantum
entanglement, the generalized GHZ states are the
most entangled quantum states.
The generalized GHZ states are the only
ones containing information at the $n$-party level.

Let $D_n$ be the set of $n$-qubit density matrices.
If $\rho \in D_n$ is an $n$-qubit density matrix,
and $j \in \{1,\ldots,n\}$ is a qubit label, we may
form an $(n-1)$-qubit reduced density matrix
$\rho_{(j)} = \tr_j \rho$
by taking the partial trace of $\rho$ over qubit $j$.
Let
\[
\PTr: D_n \to D_{n-1}^n
\]
be the map
$\rho \mapsto (\rho_{(1)},\ldots,\rho_{(n)})$
that associates to $\rho$ its $n$-tuple
of $(n-1)$-qubit reduced density matrices.
The map $\PTr$ is neither injective (one-to-one)
nor surjective (onto).
The failure of $\PTr$ to be injective means that
multiple $n$-qubit states can have the same reduced
density matrices.  States $\rho_1 \neq \rho_2$
with $\PTr(\rho_1) = \PTr(\rho_2)$
require more information for their determination
than is contained in their $(n-1)$-qubit reduced
density matrices.

Given a state $\rho$,
the set $\PTr^{-1}(\PTr(\rho))$ contains all states
with the same reduced density matrices as $\rho$.
We define a state $\rho \in D_n$ to be
\emph{determined by its reduced density matrices}
if $\PTr^{-1}(\PTr(\rho))$ contains only $\rho$,
and \emph{undetermined by its reduced density matrices}
if $\PTr^{-1}(\PTr(\rho))$ contains more than one state.

\paragraph{Main result.}
An $n$-qubit pure state $\ket{\psi}$ is undetermined
by its reduced density matrices
(among arbitrary states, pure or mixed)
if and only if $\ket{\psi}$ is LU equivalent to a generalized
$n$-qubit GHZ state.
\begin{proof}
We will show below that if a pure $n$-qubit state $\ket{\psi}$
and a mixed (non-pure) $n$-qubit state $\omega$ have the same
reduced density matrices, then there is a distinct
pure $n$-qubit state $\ket{\psi'} \neq c \ket{\psi}$
that has the same reduced density matrices as $\ket{\psi}$.
The authors showed in \cite{walcklyons08} that distinct
$n$-qubit pure states can have the same reduced density
matrices only if they are LU equivalent to
generalized $n$-qubit GHZ states.
\end{proof}

Let $\ket{\psi}$ be an $n$-qubit pure state
and let $\omega$ be an $n$-qubit mixed (non-pure) state with the
same RDMs as $\ket{\psi}$,
\[
\rho_{(j)} := \tr_j \ket{\psi} \bra{\psi} = \tr_j \omega ,
\]
where $j \in \{1,\ldots,n\}$ labels a qubit.
(In this paper, we use \emph{mixed} to mean not pure.
We use \emph{arbitrary state} to mean pure or mixed.)
For each $j$, we can Schmidt decompose $\ket{\psi}$ as
\begin{equation}
\label{psischmidt}
\ket{\psi}
 = \sum_{i=0}^1 \sqrt{q^j_i} \ket{\chi^{(j)}_i} \ket{\alpha^j_i} ,
\end{equation}
where $q^j_0 \geq q^j_1 \geq 0$ are eigenvalues of
the $(n-1)$-qubit density matrix $\rho_{(j)}$
and $\ket{\chi^{(j)}_i}$ are associated eigenvectors.

Let $\ket{\Omega}$ be a purification of $\omega$, that is, a pure state
of $n$ qubits plus an environment E, for which
\[
\omega = \tr_E \ket{\Omega} \bra{\Omega} .
\]
Because
\[
\tr_j \ket{\psi} \bra{\psi} = \tr_{j,E} \ket{\Omega} \bra{\Omega} ,
\]
we can write
\begin{equation}
\label{omegachie}
\ket{\Omega}
 = \sum_{i=0}^1 \sqrt{q^j_i} \ket{\chi^{(j)}_i} \ket{E^j_i} ,
\end{equation}
where $\ket{E^j_i}$ is a vector in
the subsystem of qubit $j$ plus the environment,
and
\begin{equation}
\label{on1}
\inprod{E^j_{i'}}{E^j_i} = \delta_{i' i} .
\end{equation}
Because
\[
\tr_{(j)} \ket{\psi} \bra{\psi} = \tr_{(j),E} \ket{\Omega} \bra{\Omega} ,
\]
where $\tr_{(j)}$ indicates a trace over all qubits \emph{except}
$j$, we can write
\begin{equation}
\label{omegaalpha}
\ket{\Omega}
 = \sum_{r=0}^1 \sqrt{q^j_r} \ket{\Omega^{(j)}_r} \ket{\alpha^j_r} ,
\end{equation}
where $\ket{\Omega^{(j)}_r}$ is a vector
in the subsystem of the environment plus all qubits except $j$,
and
\begin{equation}
\label{on2}
\inprod{\Omega^{(j)}_{r'}}{\Omega^{(j)}_r} = \delta_{r' r} .
\end{equation}

Expand
\begin{equation}
\label{bige}
\ket{E^j_i}
 = \sum_{r=0}^1 \ket{\alpha^j_r} \ket{e^j_{ir}} ,
\end{equation}
where $\ket{e^j_{ir}}$ is a vector in the
environment only, which need not be normalized nor orthogonal
to other environment vectors.

Plugging (\ref{bige}) into (\ref{omegachie}) gives
\begin{equation}
\label{omegaenv}
\ket{\Omega}
 = \sum_{i=0}^1 \sum_{r=0}^1
 \sqrt{q^j_i} \ket{\chi^{(j)}_i}
 \ket{\alpha^j_r} \ket{e^j_{ir}} .
\end{equation}
Comparing (\ref{omegaenv}) with (\ref{omegaalpha}), we see that
\[
 \sqrt{q^j_r} \ket{\Omega^{(j)}_r}
 = \sum_{i=0}^1
 \sqrt{q^j_i} \ket{\chi^{(j)}_i}
 \ket{e^j_{ir}} .
\]

The orthonormality relations (\ref{on1}) and (\ref{on2})
give conditions on the environment vectors.
\[
\sum_{r=0}^1 \inprod{e^j_{i' r}}{e^j_{i r}} = \delta_{i' i}
\]
\[
\sum_{i=0}^1 q^j_i \inprod{e^j_{i r'}}{e^j_{i r}}
 = q^j_r \delta_{r' r}
\]

\begin{lemma}
\label{env0}
If $\ket{e^j_{i i^c}} = 0$ for some qubit $j$ and some bit
$i \in \{0,1\}$ ($i^c$ is the bit complement of $i$),
and $q^j_0 q^j_1 \neq 0$, then
then $\ket{e^j_{i^c i}} = 0$.
\end{lemma}
\begin{proof}
The first orthonormality condition above
implies $\inprod{e^j_{i i}}{e^j_{i i}} = 1$,
and the second then implies $\ket{e^j_{i^c i}} = 0$.
\end{proof}

For a multi-index $I = (i_1i_2 \cdots i_n)$,
in which each $i_j \in \{0,1\}$, define
\[
\ket{I} = \ket{\alpha^1_{i_1}} \cdots \ket{\alpha^n_{i_n}} .
\]
Let
\[
\ket{\psi} = \sum_I c_I \ket{I} ,
\]
expand $\ket{\chi^{(j)}_i}$ in terms of the
$\ket{\alpha^j_i}$, and substitute into
(\ref{omegaenv}) to give
\[
\ket{\Omega}
 = \sum_{i=0}^1 \sum_{I}
 c_{i_1 \cdots i_{j-1} i i_{j+1} \cdots i_n}
 \ket{I}
 \ket{e^j_{i{i_j}}} .
\]
Equating coefficients of $\ket{I}$ with respect to
different qubits gives
\begin{equation}
\label{mainconstraint}
  c_I \ket{e^j_{i_j i_j}}
+ c_{I_j} \ket{e^j_{i_j^c i_j}}
= c_I \ket{e^k_{i_k i_k}}
+ c_{I_k} \ket{e^k_{i_k^c i_k}} ,
\end{equation}
where $I_j$ is the multi-index equal to $I$ in every
slot except $j$, where it is complemented.
\[
I_j := (i_1 \cdots i_{j-1} i_j^c i_{j+1} \cdots i_n)
\]
We call equation (\ref{mainconstraint})
the \emph{main constraint}.
It holds for every multi-index $I$ and every choice of
qubits $j$ and $k$.

\begin{lemma}
\label{twozeroonenot}
If there are qubits $j$ and $k$ and a multi-index $I$ such that
$c_I = c_{I_j} = 0$ and $c_{I_k} \neq 0$, then
$\ket{e^k_{i_k^c i_k}} = 0$.
\end{lemma}
\begin{proof}
This follows directly from the main constraint.
\end{proof}

\begin{lemma}
\label{adjacentzeros}
If $c_I = c_{I_j} = 0$ for some multi-index $I$ and some qubit $j$,
then
$\ket{e^k_{i_k^c i_k}} = 0$ for some $k$.
\end{lemma}
\begin{proof}
If $c_{I_k} \neq 0$ for some $k$, then Lemma \ref{twozeroonenot}
applies, and $\ket{e^k_{i_k^c i_k}} = 0$.
On the contrary, suppose that
$c_{I_k} = 0$ for every $k$.  In that case,
every multi-index that differs from $I$ in exactly one qubit slot
has zero coefficient.

There is at least one multi-index with a nonzero coefficient.
Choose a multi-index $I'$ with a nonzero coefficient that differs
from $I$ in a minimal number of slots, say $m$ slots.
(We know that $m$ is at least 2.)
Then all multi-indexes that differ from $I$ in $m-1$ slots
or less have zero coefficients.
Choose $k$ and $l$ to be slots in which $I'$ differs from $I$.
Then we can apply Lemma \ref{twozeroonenot}
with $I'_k$ playing the role of $I$,
$I'_{kl}$ playing the role of $I_j$,
and $I'$ playing the role of $I_k$.
We conclude that $\ket{e^k_{i_k^c i_k}} = 0$.
\end{proof}

\begin{lemma}
\label{mainlemma}
If there exist $I$ and $I'$ that differ in slot $j$
and agree in some slot $k$, such that
$c_I c_{I'} \neq 0$, and
$c_I c_{I'} - c_{I'_j} c_{I_j} \neq 0$,
then the set
\[
\{ \ket{e^j_{00}}, \ket{e^j_{01}}, \ket{e^j_{10}}, \ket{e^j_{11}} \}
\]
spans at most two dimensions.
\end{lemma}
\begin{proof}
Take $c_{I'}$ times the main constraint for $I$
minus $c_I$ times the main constraint for $I'$.
If $I$ and $I'$ differ in slot $j$, but agree in $k$, then
\begin{multline*}
c_I c_{I'} \left(
   \ket{e^j_{i_j i_j}} - \ket{e^j_{i_j^c i_j^c}} \right) \\
=
 - c_{I'} c_{I_j} \ket{e^j_{i_j^c i_j}}
 + c_I c_{I'_j} \ket{e^j_{i_j i_j^c}}
 + (c_{I'} c_{I_k} - c_{I} c_{I'_k}) \ket{e^k_{i_k^c i_k}} .
\end{multline*}
If $I$ and $I'$ differ in slot $j$, but agree in $k$, then
$I$ and $I'_j$ agree in slots $j$ and $k$, so
\[
(c_I c_{I'} - c_{I'_j} c_{I_j}) \ket{e^j_{i_j^c i_j}}
= (c_{I} c_{I'_{jk}} - c_{I'_j} c_{I_k}) \ket{e^k_{i_k^c i_k}} ,
\]
and $I_j$ and $I'$ agree in slots $j$ and $k$, so
\[
(c_{I_j} c_{I'_j} - c_{I'} c_{I}) \ket{e^j_{i_j i_j^c}}
= (c_{I_j} c_{I'_k} - c_{I'} c_{I_{jk}}) \ket{e^k_{i_k^c i_k}} .
\]
Since $c_I c_{I'} \neq 0$,
$\ket{e^j_{00}} - \ket{e^j_{11}}$ is a linear combination
of $\ket{e^j_{01}}$, $\ket{e^j_{10}}$, and $\ket{e^k_{i_k^c i_k}}$.
Since $c_I c_{I'} - c_{I'_j} c_{I_j} \neq 0$, we know that
$\ket{e^j_{01}}$, $\ket{e^j_{10}}$, and $\ket{e^k_{i_k^c i_k}}$
span one dimension.
\end{proof}

\begin{proposition}
\label{rank2}
If an $n$-qubit pure state $\ket{\psi}$ and an $n$-qubit mixed state $\omega$
have the same reduced density matrices, then $\omega$ has rank 2.
\end{proposition}
\begin{proof}
We show that $\omega$ has rank 2 by showing that at most
2 dimensions of the environment are used in $\ket{\Omega}$.
It is not possible that only one dimension of the environment
is used, because that would make $\omega$ a pure state,
contrary to our assumptions.

First we treat the case in which $q^j_0 q^j_1 = 0$ for some $j$,
which is equivalent to $\ket{\psi}$ being a product of
a single qubit state (in qubit $j$) and an $(n-1)$-qubit state.
In this case, equation (\ref{psischmidt}) becomes
\[
\ket{\psi} = \ket{\chi^{(j)}_0} \ket{\alpha^j_0} ,
\]
and equation (\ref{omegachie}) becomes
\[
\ket{\Omega}
 = \ket{\chi^{(j)}_0} \ket{E^j_0} .
\]
This expression involves only two dimensions of the environment,
so the proposition holds in this case.

From here on, we assume that $q^j_0 q^j_1 \neq 0$ for every
$j \in \{1,\ldots,n\}$.
Next suppose that $\ket{e^j_{i i^c}} = 0$ for some $j$
and some $i$.  Then, by Lemma \ref{env0}, two of the four
environment kets are zero, so the proposition holds.

Consider the case in which $c_I = 0$ for some $I$.
If there is a qubit $j$ for which $c_{I_j} = 0$ also,
then Lemma \ref{adjacentzeros} ensures that $\omega$
has rank 2.  On the other hand, if $c_{I_j} \neq 0$ for
every qubit $j$, then we can apply Lemma \ref{mainlemma}
with $j=1$, $k=2$, $I_1$ playing the role of $I$
in the Lemma and $I_2$ playing the role of $I'$.

Finally, suppose that $c_I \neq 0$ for every multi-index $I$.
If there are multi-indexes $I$ and $I'$ that disagree in some
slot $j$ and agree in some slot $k$, and satisfy
$c_I c_{I'} - c_{I'_j} c_{I_j} \neq 0$, then Lemma \ref{mainlemma}
applies, and $\omega$ has rank 2.
Otherwise, every pair of multi-indexes $I$ and $I'$ that disagree
in some slot $j$ and agree in some other slot satisfy
$c_I c_{I'} = c_{I'_j} c_{I_j}$, or equivalently,
\[
\frac{c_{i_1 \cdots i_{j-1} i_j i_{j+1} \cdots i_n}}
     {c_{i_1 \cdots i_{j-1} i_j^c i_{j+1} \cdots i_n}}
=
\frac{c_{i'_1 \cdots i'_{j-1} i_j i'_{j+1} \cdots i'_n}}
     {c_{i'_1 \cdots i'_{j-1} i_j^c i'_{j+1} \cdots i'_n}} .
\]
For $n \geq 3$, this condition
implies that $\ket{\psi}$ is
the product of a 1-qubit state and an $(n-1)$-qubit
state, contrary to our current assumptions.
\end{proof}

\begin{proposition}
If an $n$-qubit pure state $\ket{\psi}$
and an $n$-qubit mixed state $\omega$
have the same reduced density matrices, then there is a distinct
pure state $\ket{\psi'} \neq c \ket{\psi}$
that has the same reduced density matrices as $\ket{\psi}$.
\end{proposition}
\begin{proof}
Since $\omega$ has the same reduced density matrices as
$\ket{\psi}$, Proposition \ref{rank2} ensures $\omega$ has rank 2.
In fact, the matrix
\begin{equation}
\label{mixture}
(1-a) \ket{\psi} \bra{\psi} + a \omega
\end{equation}
has the same RDMs as $\ket{\psi}$ for all real $a$
that give a legitimate density matrix,
and consequently has rank at most two.
We claim that by extending $a$ beyond 1, we will reach a pure state with
the same RDMs as $\ket{\psi}$.

Let $\ket{\phi_1}$ and $\ket{\phi_2}$ be eigenvectors of $\omega$,
so that
\[
\omega = p \ket{\phi_1} \bra{\phi_1} + (1-p) \ket{\phi_2} \bra{\phi_2} .
\]
Notice that $\ket{\psi}$ is a linear combination
of $\ket{\phi_1}$ and $\ket{\phi_2}$,
\[
\ket{\psi} = c_1 \ket{\phi_1} + c_2 \ket{\phi_2} .
\]
If this were not the case, then expression (\ref{mixture}),
with $a=1/2$, say, would have rank 3.
Since $\ket{\psi}$ is in the span of
$\ket{\phi_1}$ and $\ket{\phi_2}$,
expression (\ref{mixture}) has two real eigenvalues that
sum to one for any real value of $a$.

The eigenvalues of a trace 1 Hermitian matrix
\[
\left[ \begin{array}{cc}
\frac{1}{2} + z & \bar{u} \\
u & \frac{1}{2} - z
       \end{array} \right]
\]
are
\[
\lambda = \frac{1}{2} \pm \sqrt{\abs{u}^2 + z^2} .
\]
In particular, the lowest eigenvalue is less than or equal
to the lowest diagonal element.

Now extend $\ket{\psi}$ to an orthonormal basis
$\{ \ket{\psi},\ket{\psi_2} \}$ by defining
\[
\ket{\psi_2} = \bar{c_2} \ket{\phi_1} - \bar{c_1} \ket{\phi_2} .
\]
Then,
\begin{align*}
\ket{\phi_1} &= \bar{c_1} \ket{\psi} + c_2 \ket{\psi_2} \\
\ket{\phi_2} &= \bar{c_2} \ket{\psi} - c_1 \ket{\psi_2} .
\end{align*}
The coefficient of $\ket{\psi} \bra{\psi}$
in the
$\{ \ket{\psi},\ket{\psi_2} \}$ basis
for expression
(\ref{mixture}) is
\[
1 - a + a \left[ p \abs{c_1}^2 + (1-p) \abs{c_2}^2 \right] .
\]
Since the term in square brackets is greater than zero
and less than one, this coefficient will become negative
for large enough $a$.  Hence the lowest eigenvalue of
expression (\ref{mixture}),
which is lower than either diagonal entry,
will become negative for large enough $a$.
Since $u$ and $z$ above are linear functions of $a$,
the eigenvalues are continuous functions of $a$.
Since the low eigenvalue of this matrix is
positive for $a=1$, and becomes negative for large $a$,
it must pass through zero for some value of $a$,
indicating a pure state.
It is clear that this pure state is different from
$\ket{\psi}$, since we can form $\omega$ as a mixture
of $\ket{\psi}$ and this new pure state.
\end{proof}

\paragraph{Conclusion.}
GHZ states first appeared as a way to achieve a simpler
and non-statistical rejection of local realism and local
hidden variable theories, as a way to provide a direct
example of an ``element of physical reality'' for
Einstein, Podolsky, and Rosen \cite{epr35}.
That the ``parts and whole'' view
of quantum entanglement makes a connection with
the ``rejection of local realism'' viewpoint is encouraging
and provocative.  Whether stronger connections can be
made is a subject for future work.

The authors thank the National Science Foundation
for their support of this work through NSF Award No. PHY-0555506.



\end{document}